\documentclass{article}

\usepackage[english]{babel}

\usepackage[letterpaper,top=2cm,bottom=2cm,left=2cm,right=3cm,marginparwidth=1.75cm]{geometry}

\usepackage{amsmath}
\usepackage{graphicx}
\usepackage[colorlinks=true, allcolors=blue]{hyperref}
\usepackage{graphicx}
\usepackage[utf8]{inputenc}
\usepackage{amsmath}
\usepackage{mathtools}
\usepackage{gensymb}
\usepackage{hyperref}
\hypersetup{
    colorlinks=true,
    linkcolor = black,
    anchorcolor = blue,
    citecolor = blue,
    filecolor = blue,
    urlcolor = blue
}
\graphicspath{ {Images/} }
\usepackage{epstopdf, epsfig}
\usepackage{amssymb}
\usepackage{color}

\usepackage{mathtools}
\usepackage{dcolumn}
\usepackage{bm}
\usepackage{soul}
\usepackage{float}
\usepackage{natbib}
\usepackage{authblk}

\title{Penetration and Secondary Atomization of Droplets Impacted on Wet Facemasks\thanks{This article has been accepted for publication in Physical Review Fluids. American Physical Society (APS) holds the copyright.}}%
\author[1]{Sombuddha Bagchi}
\author[2]{Saptarshi Basu}
\author[3]{Swetaprovo Chaudhuri}
\author[1,**]{Abhishek Saha}
\affil[1]{Department of Mechanical and Aerospace Engineering, University of California San Diego, La Jolla, CA-92093, USA}
\affil[2]{Department of Mechanical Engineering, Indian Institute of Science, Bengaluru, KA 560012, India}
\affil[3]{Institute for Aerospace Studies, University of Toronto, Toronto, Ontario M3H 5T6, Canada}
\affil[**]{corresponding author: asaha@eng.ucsd.edu}
\date{}                     
\begin{document}
\maketitle

\begin{abstract}
Face covering, commonly known as facemask, is considered to be one of the most effective Personal Protective Equipments (PPEs) to reduce transmissions of pathogens through respiratory droplets - both large drops and liquid aerosol particles. Face masks, not only inhibit the expulsion of such respiratory droplets from the user, but also protects the user from inhaling pathogen laden potentially harmful droplets or their dried nuclei. While the efficacies of various dry face masks have been explored in recent past, a comprehensive investigation of a wet mask is lacking. Yet, users wear masks for a long period of time and during this period, owing to respiratory droplets released through multiple respiratory events, the mask matrix becomes wet. We, herein, present an experimental study on the dynamics of sequential impacts of droplets on a masks to understand how wetness affects possible penetration and secondary atomization of the impacted droplet. Two different types of masks, hydrophobic and hydrophilic, were used in this study to evaluate the underlying physical mechanism that controls the penetration in each of them. 
\end{abstract}

\section{Introduction}

Respiratory droplets (both large drops and liquid aerosol particles) transmit various pathogens \cite{yan2018infectious, richard2020influenza, goodlow1961viability}, including SARS-Cov2 virus \cite{morawska2020time, prather2020reducing} responsible for the current Covid-19 pandemic which has taken close to 4M lives and infected close to 182M people worldwide at the time of writing this article \cite{world2020transmission}. Such droplets, when exhaled through the nasal and oral cavities of the infected individual during respiratory events such as speaking, singing, coughing, sneezing etc, transport and carry the pathogens \cite{duguid1946size, Abkarian25237, Yang_Virus_2020, Basu-levitator-2020}. Due their initial momentum and surrounding gas-phase flow, these droplets can travel long distances while they undergo evaporation (or condensation) and precipitation \cite{bourouiba2014violent, Dbouk_droplet, chaudhuri2020modeling,  Ng_Growth_repiratory,Chong_extended_life, prather2020reducing}. These airborne droplets, if inhaled by the healthy individuals, can deposit in the respiratory tracts\cite{liu2020aerodynamic} causing infection.
To curb the spread of diseases which transmit through respiratory droplets, one of the active modes of protection often suggested is the use of proper face-covering, commonly know as masks \cite{esposito2020universal,leung2020respiratory}. Good masking practice can provide two levels of securities. On one hand, the mask can block the respiratory droplets (and the pathogens) during exhalation protecting others, and on the other hand, it also protects the wearer by blocking virus laden aerosols/droplets during inhalation \citep{li2020mask,eikenberry2020mask}.
\citet{bandiera2020face} estimated reduction in exposure to these droplets almost by 10000 times when standing very close (within 0.5m) to a masked individual compared to standing far (2m) from an unmasked individual. Using their susceptible-exposed-infectious-recovered (SEIR) model, \citet{chaudhuri2020analyzing} showed that while cough droplets with initial diameter of 10-50$\mu m$ have a very high potential of infection (high reproduction number, $R_{0,c}$), universal masking which blocks all droplets sizes above 5$\mu m$, could result in significant decrease in infection transmission ($R_{0,c} \leq 1$). 

While face masks can be effective in preventing the transmission of the pathogens, not all forms of masks or face-covering offer proper protection. Thus, it is critical to assess the performance of various masks or mask materials. One of the first studies on the efficacy of face masks was performed by \citet{weaver1919droplet} in 1919. 
The necessity of face covering became critical during early phases of Covid-19 pandemic, and many places still mandate use of mask for enhanced protections. However, it is worth noting that, although N95 masks was deemed to be the most effective in blocking the transmission, the increased demand and cost led to usage of many alternate forms of masks in both developed and developing countries. Thus, a great amount of research effort around the world over last 12-16 months have focused on understanding the fundamentals of the various masks and assessing their efficacy. \citet{fischer2020low} developed a cost-effective measurement
technique to assess the performance of various masks by comparing the number and the volume of the ejected droplets for different masks. In particular, they showed that while cotton cloth masks provide similar efficacy to the surgical masks, the other alternatives, such as neck gaiters or bandanas provide minimal protection. \citet{ho2020medical} also showed that cotton cloth masks can be used as a substitute for medical masks by healthy individuals without significantly compromising the risk potential. Numerical simulations by \citet{Dbouk_droplet} demonstrated that respiratory droplets, if escaped through the masks, can travel to a long distance. Using Mie scattering, \citet{Verma_visualization} showed that the travel distance of the cough jets (gaseous volume exhaled during coughing) reduces significantly with masks. However, the side-leakage can still cause many aerosolized droplets to escape. \citet{asadi2020efficacy} studied the emissions of microscale particles during respiratory events and showed that the usage of proper masks can reduce them significantly. 

One of the most common mask used in both medical and social settings is surgical mask. Although there are variety of surgical masks available in market, most of them contains three layers. In our recent exploration \citep{sharma2021secondary}, we studied the possibility of secondary atomization of large respiratory droplets through these surgical masks. By observing a single large ($>200 \mu m$) droplet impacting on single, and multi-layer masks, we showed that, when the impact velocity is beyond a critical limit, due to hydrodynamic focusing the impacted droplet can penetrate the mask matrix producing a large number of smaller droplets through secondary atomization. {Such penetration and secondary atomization have also been reported for droplets, impacted on dry uniform rigid meshes}\citep{kooij2019sprays}{. The atomization process, here, is driven by formation and subsequent instability of liquid ligaments during the penetration process}\cite{villermaux2007fragmentation}.  Furthermore, based on the work by \citet{Sahu_porous_2012} on porous media, we formulated a scaling law for the critical condition required for penetration, which confirmed that the critical velocity for penetration increases for thicker or multi-layer masks, decreasing the propensity of atomization \cite{sharma2021secondary}. The study also demonstrated that a large number of the atomized droplets are smaller than $100 \mu m$, a range often considered to be aerosol due to their long airborne lifetime \cite{prather2020reducing}. 
In a separate study, \citet{melayil2021wetting} found that excessive hydrophobic coating on the inner layer can also lead to bouncing of impacted respiratory droplet, leading to generation of smaller daughter droplets.

While the previous studies have provided a great insight on mask efficacy and the potential of atomization, most of the studies considered a dry mask matrix as the medium. In reality, we use mask for a prolonged period where the matrix is exposed to countless respiratory events including breathing, talking, coughing, sneezing etc. The respiratory droplets emanating during these events impact on the mask matrix, and in the process they wet the matrix irrespective of the degree of blockage. Furthermore, in humid conditions, wetting of masks can also occur through condensation of exhaled moist air \cite{Ng_Growth_repiratory} and sweating. Thus, most of the time, we essentially use a wet mask, whose wetness progressively increases with time and frequency of respiratory events. Although evaporation can lead to drying of the mask matrix, the complete dryness is never achieved. It is then critical to understand the fate of the respiratory droplets when they impact the wet mask matrix. Recognizing its criticality and fundamental importance, in this study we explore the effect of wetness on the impact of respiratory droplet on mask matrix. In particular, we set to answer two questions. (1) How does the critical condition for penetration and secondary atomization change for a wet mask as its wetness progressively increases? (2) How does hydrophobic vs hydrophilic mask surfaces perform differently once the matrix is wet? To find answers to these questions, we conducted a sequential droplet impact experiments on masks, and hence simulated the effect of multiple respiratory events which progressively increases the wetness of the mask. We will show that the morphology of impact changes as the mask matrix becomes wet and the degree of wetness affects the hydrophobic and hydrophilic masks differently.

\begin{table}[t]
    \centering
    \begin{tabular}{ |p{3cm}||p{3cm}|p{3cm}|p{3cm}| }
    \hline
    &\textbf{Mask A}&\textbf{Mask B}&\textbf{Mask C}  \\
    \hline
    \textbf{Description} & Surgical mask (1-layer) & Cloth mask(Type-I) & Cloth mask(Type-II) \\
    \hline
    {\textbf{Characteristics}} & {Hydrophobic} & {Hydrophilic} & {Hydrophilic} \\
    \hline
    \textbf{Source} & Allinon Advance Technology & American Mask Project & Tultex Face Mask \\
    \hline
    \textbf{Image} & \includegraphics[width=0.9\linewidth]{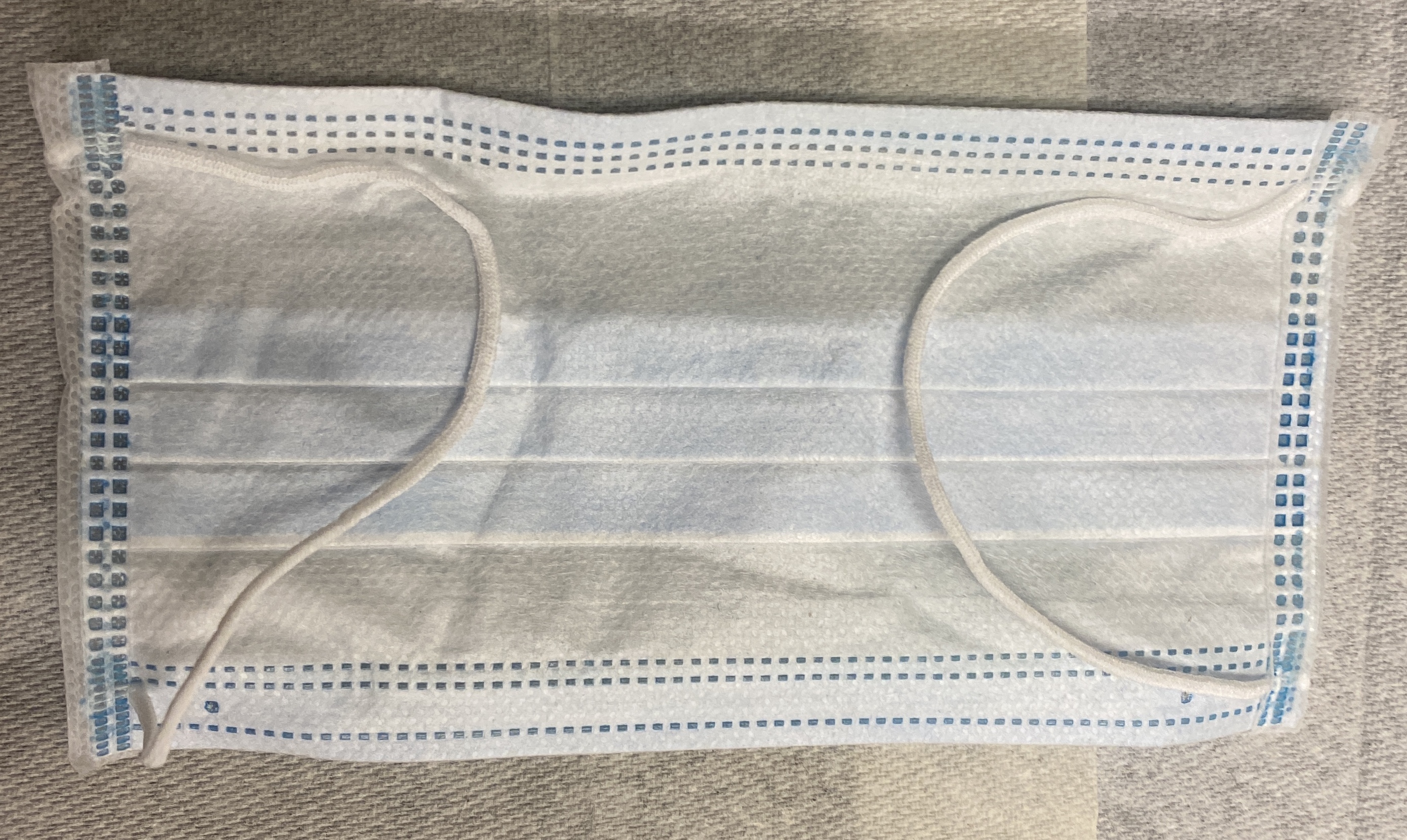} & \includegraphics[width=0.9\linewidth]{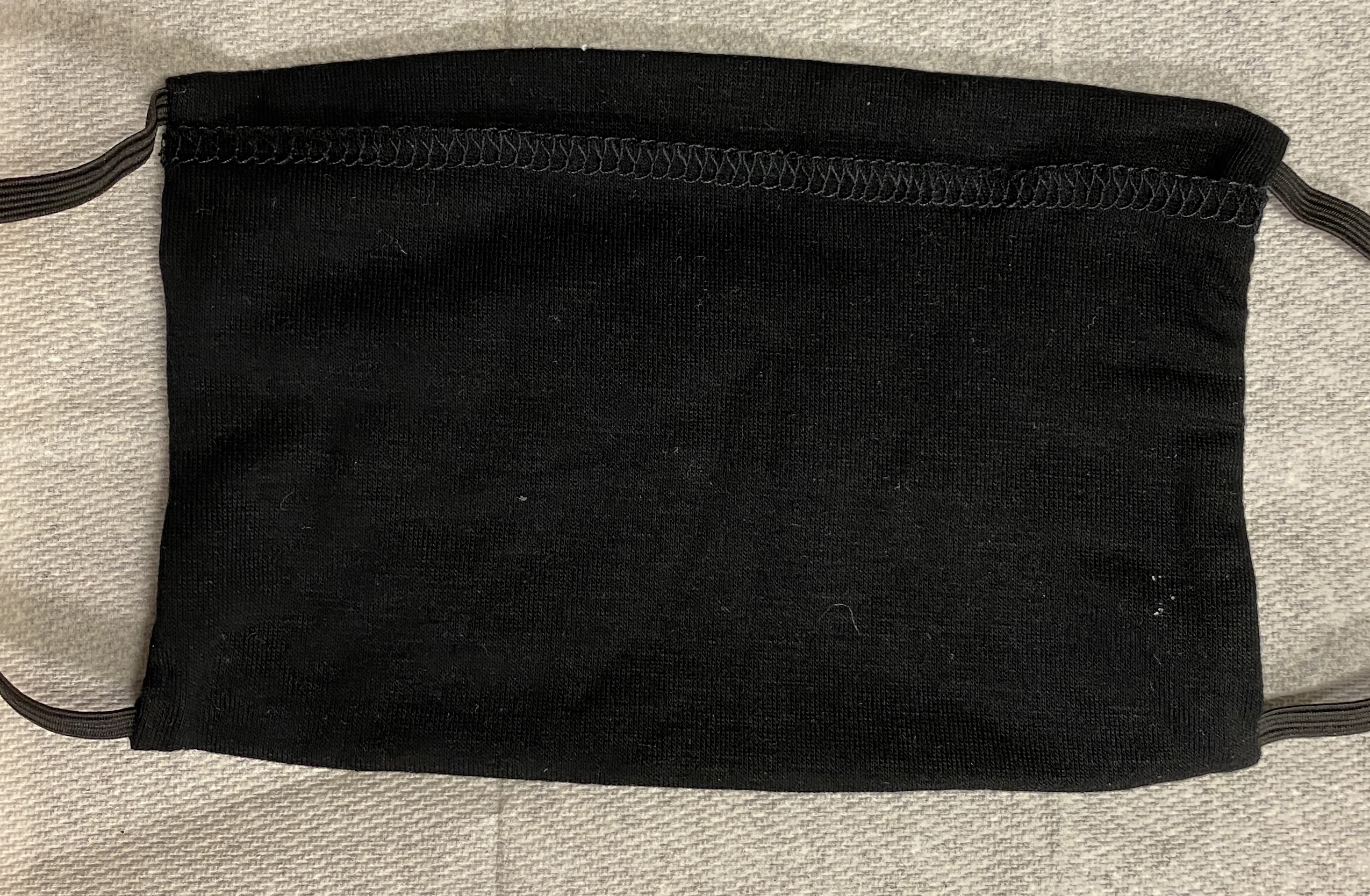} & \includegraphics[width=0.9\linewidth]{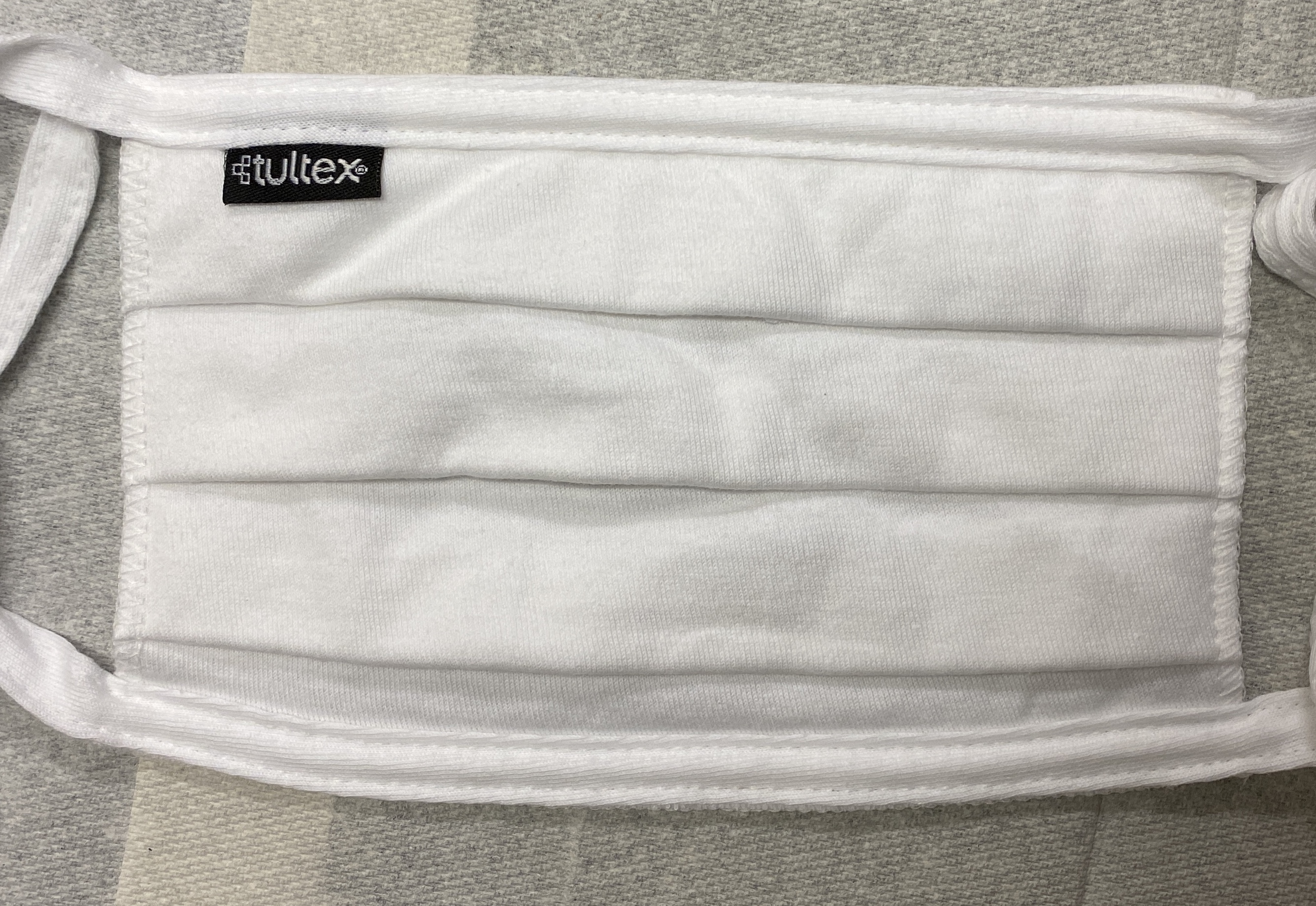} \\
    \hline
    \textbf{Thickness, $t_m$ ($\mu$m)} & $175\pm1$  & $394\pm26$ & $426\pm31$\\
    \hline
    \textbf{Pore Size $\epsilon$ ($\mu$m)} & $37.7\pm1.9$ & $140.7\pm24$ & $172.4\pm13.8$ \\
    \hline
    \end{tabular}
    \caption{Details of the three different masks used in the current study}
    \label{tab:mask}
\end{table}

\section{Experimental details}
In our experiment, the droplets were generated by a syringe pump, which slowly pushes liquid through a vertically placed needle. When the droplet at the tip becomes larger than a critical size, due to gravity it detaches from the needle and lands on the strip of mask material, placed directly below the needle. The masks were tightly held by two clamps such that they maintain a horizontal surface. The impact velocity was varied by changing the vertical distance between the mask and the needle. The impact process was captured by a high-speed camera (Phantom V710) connected to a 24-70mm lens and lens-extenders, which provides image resolution of $\sim$47 px/mm. The events were recorded at 4000fps with 600px$\times$800px window size, while a uniform bright background was created using a LED lamp with a diffused screen. 
To simulate series of respiratory events, droplets were generated at regular intervals without changing the impacted surface and the impact process was captured. To maintain ``quasi-steadiness'', a significant interval (about 30s) was maintained between impact of two subsequent droplets. As we move in the sequence of the impacts, the wetness of the mask matrix progressively increases. Although the droplets in one sequence is expected to land on the same spot of the mask, sometimes, particularly for higher impact velocities, there could be large lateral shift in the droplets. To be consistent in our analysis, we only considered the impacts where the droplets landed within $\pm2D$ of the initial location. Here $D$ is the droplet diameter. Each velocity condition was repeated several times using different section of the mask to achieve statistical convergence.

\begin{figure}[h]
\centering
\includegraphics[width=0.7\linewidth]{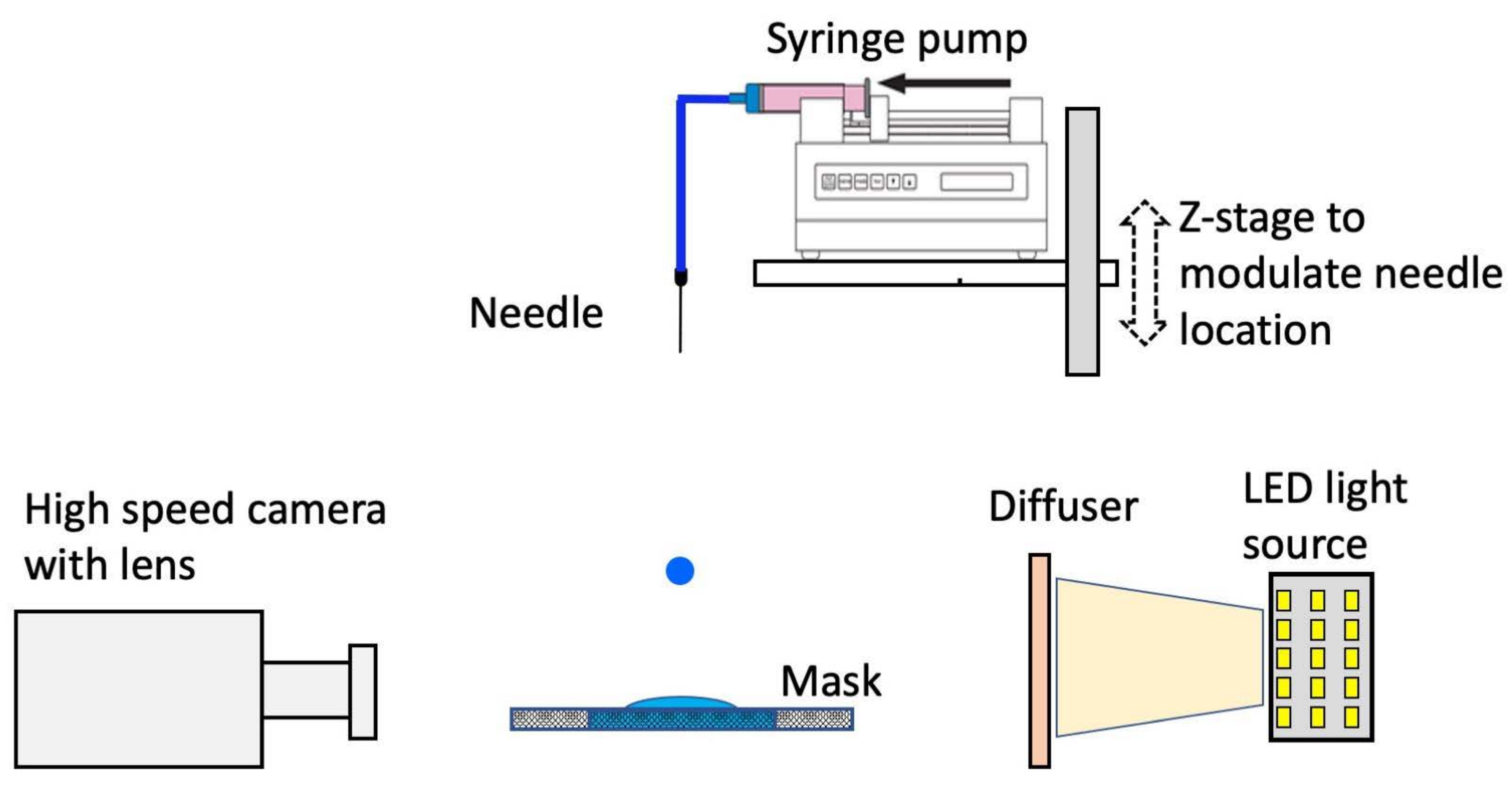}
\caption{{Primary components of the experimental setup. The components are not drawn to scale.}}
\label{fig:exp_setup}
\end{figure}

We used three different single-layer masks for these experiments. Mask A is the inner-layer of a surgical mask; Mask B is a inner layer of a cloth mask, while Mask C refers to a thicker, but single layer cloth mask. The detailed information of these masks and the sources are provided in Table \ref{tab:mask}. The average thickness ($t_m$) of the masks were measured using precision micrometers, while average pore size ($\epsilon$) was measured using backlit microscopic images (details in supplementary material). Here, pore sizes are defined as the mean equivalent diameters of the white space in the images. It is to be noted that the surgical mask, in general, is thinner but it contains smaller pore sizes, while the cloth masks are thicker but has larger pore sizes. We also attempted to measure the contact angle by placing and imaging a sessile droplet on the mask. Masks A, the surgical mask, has a contact angle of $\sim 106^{\circ}\pm10^{\circ}$, and can be treated as hydrophobic. On the other hand, once a droplet is placed on the cloth masks (B and C), they immediately absorbs the liquid showing strong affinity to water. The measurement of contact angle was not possible for these two hydrophilic masks.  

{We note that the respiratory fluids contains small quantity of salts, proteins and surfactant, all dissolved in large proportion of ($>90\%$) water} \cite{gould2001expression,effros2002dilution,Vejerano2018,balusamy2021lifetime}. 
{While their exact concentrations vary from subject-to-subject, for scientific investigations often a surrogate respiratory fluid is used, which possesses similar thermo-physical and chemical characteristics of respiratory liquids} \cite{Vejerano2018}. {In our previous study} \cite{sharma2021secondary} {with droplet impact on dry masks, we have shown that the critical conditions for the penetration does not change significantly between DI water and surrogate respiratory droplets. It is particularly because the majority ($>$90\%) of the respiratory fluid is water and as such, the thermo-physical properties are very close to those of water. Therefore, in this study, we only used pure (DI) water (density: $\rho_l\approx1000$kg/m$^3$; dynamic viscosity: $\mu_l\approx1$cP; surface tension: $\sigma\approx72$N/m) as our test liquid. But we note that, in rare occasions the respiratory fluid may contain a high percentage of mucus, which will introduce viscoelastic effects} \cite{lai2009micro}, {and as such the penetration process for these droplets may have some differences.}

{The nominal droplet diameter used for this study is about 2$mm$ for several reasons. The smaller droplets suffer from larger lateral shifts during experiments, particularly for higher velocity impacts since it requires a greater distance between the needle and the impact location. Thus, to minimize the uncertainty in the experiments, we used large droplets.
Although only a small number of large droplets (1-2 $mm$) are ejected during respiratory events} \citep{duguid1946size,xie2007far,xie2009exhaled,mittal2020flow} {(see supplementary material), these large droplets carry almost 80\% of the liquid volume}\cite{duguid1946size,sharma2021secondary} {ejected (see supplementary material). Since the number of pathogens (bacteria / virus) carried, depends on the droplet volume, the relevance of these large droplets to disease transmission is critical. Furthermore, the large droplets (mm-size), when ejected without restrictions of face covering, settle on the ground rather quickly due to gravity. However, as shown in our previous study} \cite{sharma2021secondary} {and will be shown in the present study, these large droplets may lead to the fragmentation and generation of numerous smaller daughter droplets with significant translational velocity, when inefficient masks are used. These small droplets can remain aerosolized for a long duration enhancing their contribution towards disease transmission. 
Finally, we also note that in our previous study} \cite{sharma2021secondary} {with dry masks, we have shown that the penetration characteristics do not change significantly for a range of droplet diameters (250$\mu m$-1.2 mm). }


\section{Results}
\subsection{Penetration Morphology}

\begin{figure}[]
\centering
\includegraphics[width=0.75\linewidth]{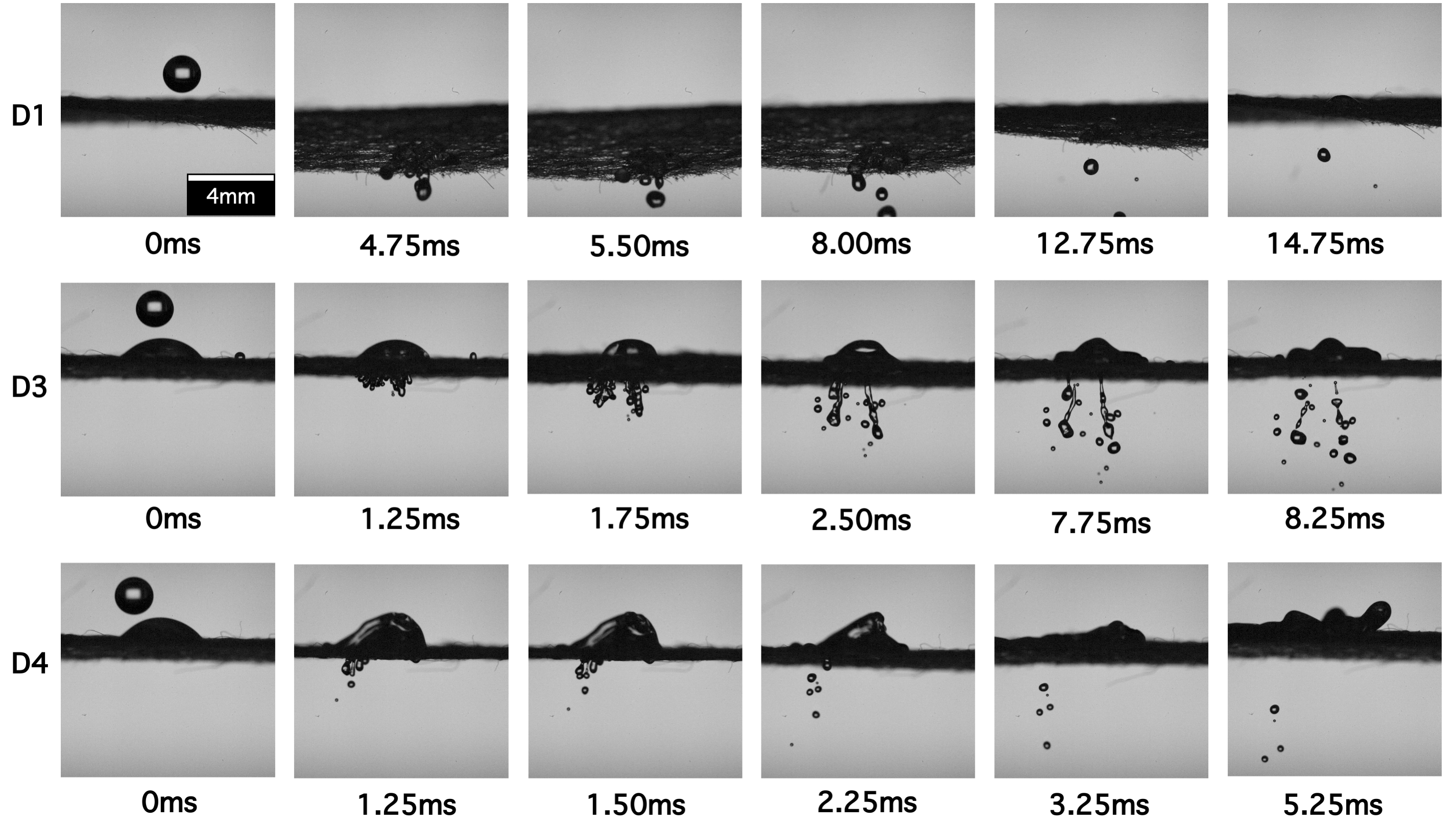}
\caption{High speed image sequence of droplet impact on Mask A for $1^{st}$ (D1), $3^{rd}$ (D3) and $4^{th}$ (D4) impacts. Recording rate is 4000fps (frames per second). The impact velocity, $U$, is 2.7m/s.}
\label{fig:HS_surgical}
\end{figure}

In our previous study \cite{sharma2021secondary} we illustrated the impact dynamics on dry mask, and showed that there exists a critical velocity ($U_{cr})$ beyond which impacts result in penetration and secondary atomization. Here, we are interested in such dynamics for subsequent impacts, i.e. when the mask surface is already wet. 
Note that we expect the dynamics to change as the wetness of the mask increases due to sequential droplet impacts. To denote a particular impact in the sequence, we will use the notation ``D\#'', i.e. D1 for the first droplet, D2 for the second droplet, etc. 
For a impact velocity ($U\approx2.7m/s$), Fig. \ref{fig:HS_surgical} shows three of
these impacts on Mask A, all of which shows penetration. The top row (D1) demonstrates the impact of the first droplet where the mask was dry. We see that this impact resulted in partial penetration, in that part of the droplet volume penetrated the mask matrix and formed smaller secondary droplets. In general, rest of the volume of liquid remains in the mask matrix. 
Subsequent droplets (i.e. droplets 2, 3, ...etc) then impact this wetted media. The impact of $3^{rd}$ and $4^{th}$ droplets are shown in the second and the third rows (D3 and D4) of Fig. \ref{fig:HS_surgical}, both of which also results in penetration. 
However, we note one key change when the mask is wet. Mask A being hydrophobic in nature, its media has limited affinity for water molecules, and hence, low absorptivity. Thus, when the remnant volume retained by the mask matrix beyond penetration, increases, they leave the mask matrix and forms small sessile droplets, often called ``beads'' on the masks surface. The first images for D3 and D4 in Fig. \ref{fig:HS_surgical}), show such ``beads'' formed by the remnant liquid from previous impacts. As shown in the figures, the subsequent droplets impact on these ``beads'' which provide additional resistance for the impacted droplets against possible penetration. In rare occasions these impacts on deposited liquid can lead to bouncing in which part of the impacted droplet gets separated. Such bouncing at low impact velocity has been shown for impact on liquid films in other studies \cite{tang-Langmuir-2018, tang-POF-2019}. For the range of velocity studied in this work, such bouncing happened in rare occasions ($<1\%$) and as such neglected for the statistical analysis presented in next sections. 
Following this discussion, it is clear that the impacted droplet volume can contribute to one of the three possible outcomes, (a) penetrated volume ($V_p)$, the volume which penetrated and travelled downstream of the mask in form of secondary droplets, (b) absorbed volume ($V_a$), the volume which is absorbed by the mask's porous matrix and (c) reflected volume ($V_r$), the volume which remained on the mask surface as ``beads''. From volume conservation, one can write $V_p+V_a+V_r=V$, where $V$ is the volume of impacting droplet. By analyzing the high speed image sequence, we can quantify $V_p$ and $V_r$ and from the volume conservation, subsequently,  we can evaluate $V_a$. Although we will discuss these in details in the next section, from visual inspection of Fig. \ref{fig:HS_surgical}, we can expect $V_r$ to be relatively large for hydrophobic Mask A. 

\begin{figure}[]
\centering
\includegraphics[width=0.75\linewidth]{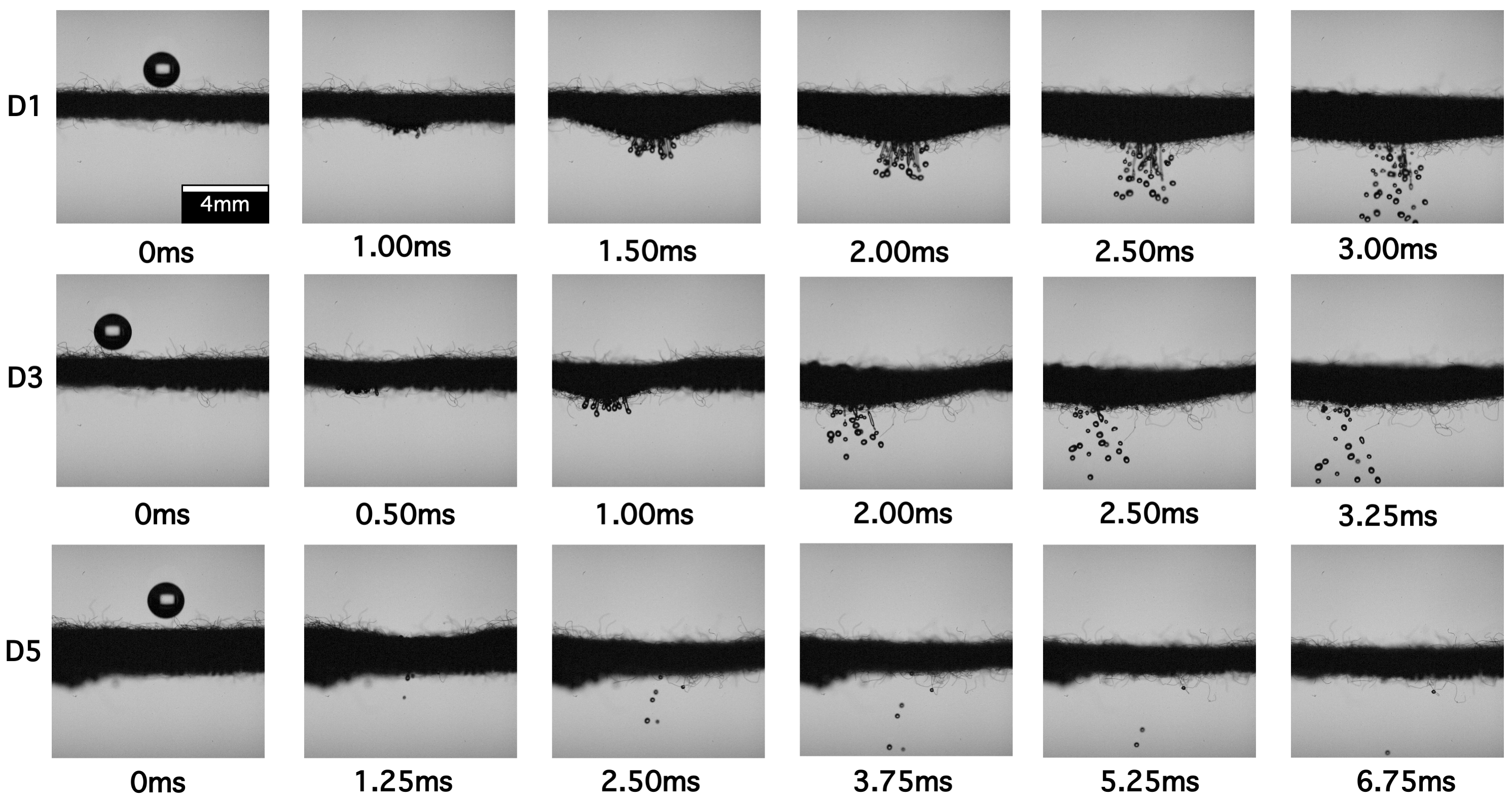}
\caption{High speed image sequence of droplet impact on Mask B for $1^{st}$ (D1), $3^{rd}$ (D3) and $5^{th}$ (D5) impacts. Recording rate is 4000fps. The impact velocity, $U$, is 3.3m/s.}
\label{fig:HS_black}
\end{figure}

In Fig. \ref{fig:HS_black}, we showed similar impacts on hydrophilic mask (Mask B) for impact velocity $U\approx3.3 m/s$, all of which also exhibit penetration. The critical morphological distinction between the impact process in Mask A and Mask B is that the latter, being a hydrophilic cloth mask, has superior absorptivity for water and as such its porous matrix can hold larger volume of water due to molecular affinity and capillary attraction \cite{Fei_Fiber_water}. Thus, we don't see formation of beads when the mask is wet, resulting is negligible $V_r$ for cloth masks. Similar dynamics is also observed for Mask C. For high-speed snapshots of the impact processes, please refer to the supplementary material.

\subsection{Penetration Dynamics}
\label{regime_map}
Now that we have visually described the penetration process observed for two types of masks studied in this work, in this section, we will discuss each of their behaviors quantitatively. It is to be noted that the microscopic description of the penetration and atomization requires a detailed investigation of masks morphology, and properties of the threads used in those. However, the primary goal of this study is to provide a generalized insight on the effect of wetness on the penetration dynamics. Hence, we will use phenomenological description to elucidate the fluid mechanical aspect of the problem. 

\begin{figure}[]
\centering
\includegraphics[width=\linewidth]{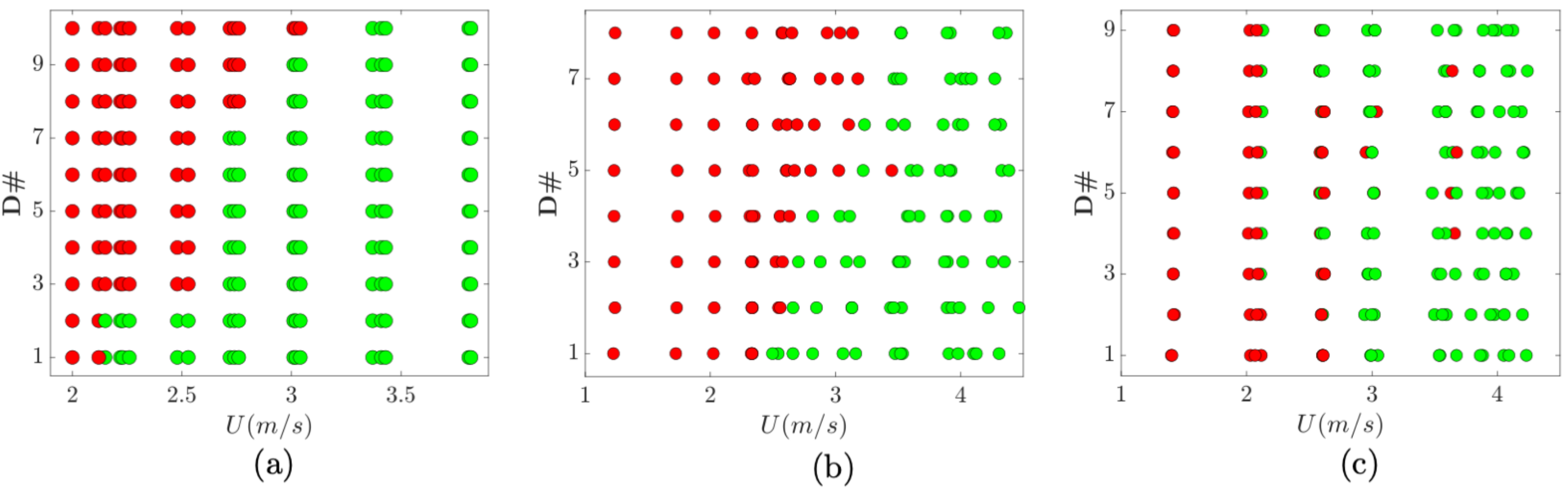}
\caption{Regime map of three different masks (a) Mask A (b) Mask B, and (c) Mask C. The green and red dots denote penetration and no-penetration, respectively.}
\label{fig:regime}
\end{figure}

\subsubsection{Hydrophobic Mask}
As stated in the introduction, in our previous study \cite{sharma2021secondary} we highlighted that for dry masks, penetration occurs beyond a critical velocity, $U_{cr}$. In the current study we explore the effect of wet-mask on such a critical velocity. This is best illustrated in a regime map (Fig. \ref{fig:regime}) where penetration-vs-no-penetration is identified as function of the impact velocity, $U$ and impacted droplet number, D\#.

In Fig. \ref{fig:regime}a, we note that under dry condition mask (D1), for mask A (hydrophobic surgical mask), the $U_{cr}$ is about $2.1m/s$, and for lower impact velocities penetration is not achieved. Working with droplet impact on dry porous media, \citet{Sahu_porous_2012} showed that the penetration occurs when the kinetic energy ($E_k$) of the liquid in pores, exceeds the viscous loss ($E_d$). In their analysis, they showed that the kinetic energy of the liquid in pores can be estimated as $E_k\sim\rho_l(U_p)^2D^3$, where $U_p\sim U(D/\epsilon)$ is the velocity of the liquid in pores. Here, $\epsilon$ is the average pore size and $D$ is the impacted droplet diameter. Since the former is smaller than the latter (see Table \ref{tab:mask}) we see that $U_p>U$, a phenomenon known as hydrodynamic focusing \cite{weickgenannt2011nonisothermal,lembach2010drop}. On the other hand, the shear stress in each pores can be expressed as $\tau_s\sim \mu_l (U_p/\epsilon)$ and thus, the total viscous dissipation becomes $E_d\sim \tau_s \epsilon D t_m N_d$. Here, $t_m$ is the average thickness of the porous media. $N_d$ is the number of active pores in dry mask during penetration process and from volume conservation, they showed $N_d\sim (D/\epsilon)^3$. Finally, we arrive at the scaling for the viscous dissipation as $E_d\sim(\mu_l U D^5 t_m)/\epsilon^4$. Since the impact Weber number, $We=\rho_l U^2 D/\sigma$, is high ($>100$), the surface tension effects are generally neglected in such scaling analysis for the penetration process. Using the above scaling and recognizing that the condition for penetration is $E_k\gg E_d$, in our previous work \cite{sharma2021secondary} with dry mask, we expressed the critical condition as 
\begin{equation}
    Re(\epsilon/t_m)\gg1
    \label{eq:Re_dry}
\end{equation}
Here, Reynolds number is defined as $Re=\rho_lU\epsilon/\mu_l$. $\rho_l$, $\mu_l$ and $\sigma$ are liquid density, dynamic viscosity and surface tension, respectively. 

We note that Eq. \ref{eq:Re_dry} is derived for dry masks and once the mask matrix is wet, the transitional condition is expected to change. This can be precisely observed in the regime map for Mask A (Fig. \ref{fig:regime}a), where we notice that subsequent impact (D3 onward) on wetted masks, the penetration occurs progressively at higher velocities. In fact, for $10^{th}$ impact (D10), $U_{cr}$ becomes about $3.4m/s$, which is more than 50\% increase over the value ($2.1 m/s$) for dry mask. Next, we will analyze the penetration process in details to understand why the wetted hydrophobic mask renders the penetration process difficult and $U_{cr}$ increases progressively with D\#. 

\begin{figure}[]
\centering
\includegraphics[width=0.8\linewidth]{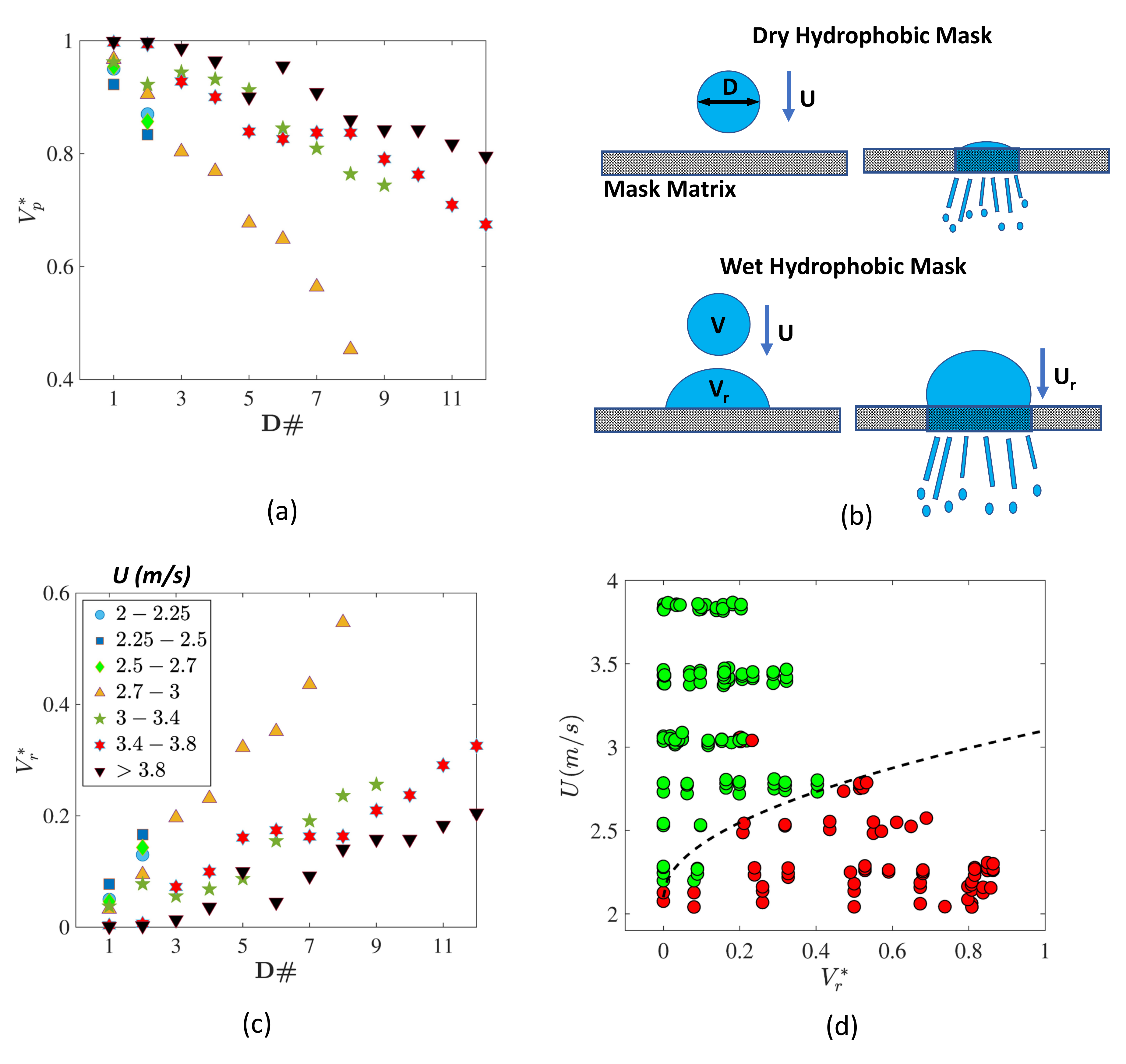}
\caption{Dynamics of droplet impact on hydrophobic mask (Mask A). (a) Normalized volume of liquid penetrated, $V^*_p=V_r/V$ through the Mask A. Various impact velocities $U$ (in m/s) are shown as different symbols, as noted in legend in (c). The impact velocities $U(m/s)$ are reported in the legend with different symbols. (b) Schematic of droplet impact on dry vs wet hydrophobic masks (not drawn to scale). (c) Normalized volume reflected (volume of ``beads''), $V^*_r=V_r/V$, after each impact reported in (a). (d) Modified regime map as function of impact velocity $U$ and normalized reflected volume, $V^*_r$. Green dots: penetration. Red dots: no penetration. the dashed line is scaling, $U_{cr}=U_0+A_r(V^*_r)^{1/2}$. Here, $U_0=2.1 m/s$ and scaling constant $A_r=1m/s$.}
\label{fig:hydrophobic_all}
\end{figure}

We recognize that, while the regime map provides a clear understanding of when we expect penetration to occur, it does not necessarily illustrate how much liquid has penetrated through the mask in the form of secondary atomization or daughter droplets. In contrast, the quantity volume penetrated, $V_p$, provides better insight. Analyzing the high speed image sequences and by tracing the individual daughter droplets produced after each impact using ImageJ image processing software \citep{schneider2012nih}, we evaluated the normalized penetration volume, defined as $V^*_p=V_p/V$ and plotted as function of D\# for various impact speeds as shown in Fig. \ref{fig:hydrophobic_all}a for impact on hydrophobic mask. Here, $V$ is volume of the original droplet and $V_p$ is the total volume of the secondary droplets produced after each impact. For the impact on dry masks (D1), we observe that $V^*_p$ is close to unity for all values of $U$, signifying that most of the liquid has penetrated. However, once wetted (D2 onwards), the penetrated volume decreases almost monotonically. Since the impacted droplet diameter and hence, the volume ($V$) remains roughly constant during each impact, this trend suggests that the remnant volume will increase with D\#. On the other hand, as pointed out before, hydrophobic porous matrix, such as Mask A, cannot absorb large volume of liquid. Thus, the remnant liquid after partial penetration during the impacts, accumulates on the top of the mask in form of ``beads''. The schematic of the impact process on a dry-vs-wet hydrophobic mask in shown in Fig. \ref{fig:hydrophobic_all}b. Analyzing the high speed images and assuming that the ``beads'' are axisymmetric in nature, we can evaluate the normalized reflected volume $V^*_r=V_r/V$ after each impact as shown in Fig. \ref{fig:hydrophobic_all}. We clearly see an almost monotonic increase in ${V^*_r}/V$ with D\#. From the values reported in Figs. \ref{fig:hydrophobic_all}a and c, we see that $V^*_p+V^*_r\approx 1$ for any given $U$ and D\#. This confirms our hypothesis that $V^*_a\approx 0$, i.e. the matrix of hydrophobic mask does not retain large quantity of liquid.

As illustrated in the schematic (Fig. \ref{fig:hydrophobic_all}b), once beads are formed the subsequent droplets land on this accumulated liquid and pushes it through the mask matrix underneath. Due to increased mass, the average effective velocity, $U_r$, will reduce from the original impact velocity of the droplet, $U$. Such reduction in impact velocity and hence lack in kinetic energy will render penetration more difficult. From energy balance, we get the scaling relation $U_r \sim U(V/V_r)^{1/2}$. Substituting this in the scaling analysis presented in Eq. \ref{eq:Re_dry}, we find a modified scaling for $U$ at critical condition for wet hydrophobic masks,
 \begin{equation}
     U_{cr}\sim A_0(V^*_r)^{1/2}.
     \label{eq:scaling_hydphobic}
 \end{equation}
The factor $A_0=(\mu_l t_m) /(\rho_l \epsilon^2$) is fixed for a given mask, liquid and droplet size. Equation \ref{eq:scaling_hydphobic} suggests that the critical velocity, $U_{cr}$ required for penetration increases with $V_r$, which, on the other hand, increases with D\# as shown in Fig. \ref{fig:hydrophobic_all}c. Due to the combined effect, penetration becomes more difficult with increased wetness as shown in regime map (Fig. \ref{fig:regime}a). It is worth noting that a loss in effective kinetic energy has also been reported in previous studies for droplet impact on liquid pools supported by either rigid \cite{tran2013air, tang-SoftMatter-2016, saha-JFM-2019} and soft surfaces \cite{Shin-SoftMatter-2020}. 

Guided by the scaling relation in Eq. \ref{eq:scaling_hydphobic}, Next, we present a modified regime map for the hydrophobic mask (Mask A), where we plot the impact outcomes (penetration vs no-penetration) as function of $V^*_r$ and $U$ as shown in Fig. \ref{fig:hydrophobic_all}d. We note that, apart from few scatter points, the transition from penetration (green dot) to no penetration shift to a higher $U$ as $V^*_r$ increases, which qualitatively supports the scaling in Eq. \ref{eq:scaling_hydphobic}. To check quantitatively, we added the scaling as reference line, $U_{cr}=U_0+A_r(V^*_r)^{1/2}$ (shown with dashed line) in the same figure. The term, $U_0=2.1m/s$ is the critical velocity for dry mask, was added to the scaling to ensure that as $(V^*_r)\rightarrow 0$ for dry mask, we get $U_{cr}=U_0$. With scaling parameter, $A_r=1m/s$, we observe a good agreement between data and scaling, in that the transition between penetration (green dots) to no-penetration (red dots) are reasonably well captured by the line.

\subsubsection{Hydrophilic Masks}
The regime maps for two hydrophilic masks (Mask B and C) are shown in Fig. \ref{fig:regime}b and c. 
Mask B, which has a $U_{cr}$ of $2.5m/s$ under dry condition (D1), shows an consistent increase in $U_{cr}$ for the subsequent impacts. For $10^{th}$ droplet (D10), $U_{cr}$ becomes almost $3.5m/s$. On the other hand, for Mask C, at the dry condition $U_{cr}$ is roughly $2.6m/s$. As this masks gets wet during subsequent droplet impacts, $U_{cr}$ slightly decreases to 2.1m/s for the 3rd droplet (D3) and then remains unaltered for subsequent impacts. There are some uncertainties in determining the $U_{cr}$ for this masks, as we note overlap of penetration (green dots) and no penetration (red dots) outcomes around the transitional boundary. Nevertheless, unlike Mask B, we do not see a clear sharp increase in $U_{cr}$ with D\# for mask C. Next, we will show the underlying controlling mechanism for hydrophilic masks (B and C) are indeed same although they display different qualitative behavior in the regime maps. We will also demonstrate that the penetration dynamics for these hydrophilic masks are different than that for hydrophobic masks, described in the previous section, although the regime maps from hydrophobic Mask A and hydrophilic Mask B look qualitatively similar.

\begin{figure}[]
\centering
\includegraphics[width=0.75\linewidth]{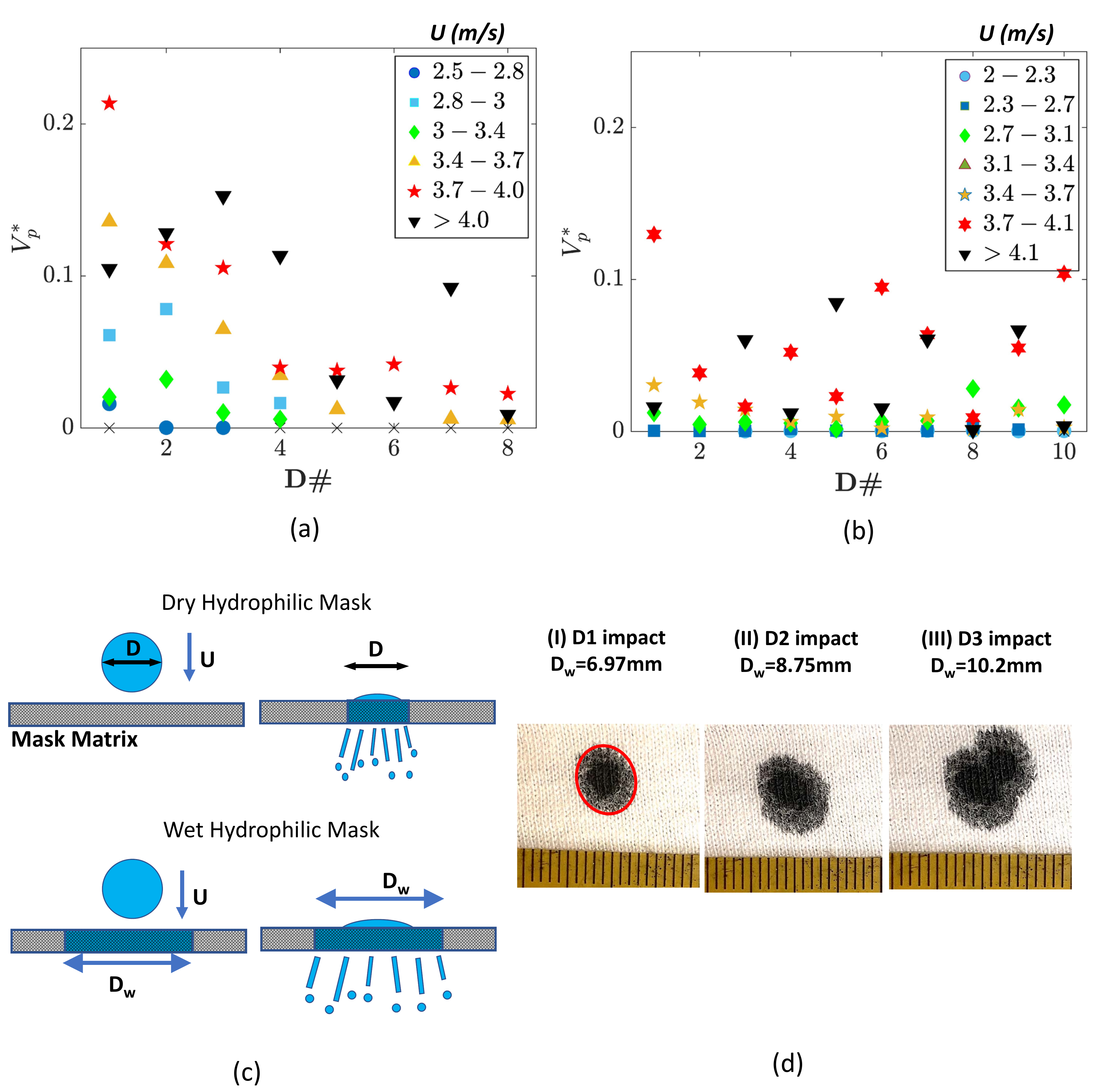}
\caption{Dynamics of droplet impact on hydrophilic mask (Mask B and C). (a) and (b) Normalized volume of liquid penetrated, $V^*_{p}=V_p/V$, through (a) Mask B and (b) Mask C. Different symbols are for various impact velocity $U(m/s)$, as shown in legends. (c) Schematic of droplet impact on dry vs wet hydrophilic masks (not drawn to scale). (d) Spread of droplets on hydrophilic mask with increase in the number of droplet impacts.}
\label{fig:hydrophilic_all}
\end{figure}

Since cloth masks are hydrophilic and has strong affinity of water, a major portion of the impacted droplet volume is absorbed by the fibres of the mask, leading to very different dynamics compared to hydrophobic masks (Mask A). This is also reflected in Figs. \ref{fig:hydrophilic_all}a and b, where we plotted the $V^*_p$ for the hydrophilic masks. We note that at dry condition (D1),  $V^*_p\ll 1$, while for Mask A, we observed $V^*_P\approx 1$ at D1 (Fig. \ref{fig:hydrophobic_all}a). 

Furthermore, unlike Mask A, the hydrophilic cloth masks (Masks B and C) do not exhibit ``beading'',  as the porous matrix can absorb water. In fact, as more and more droplets are impacted the wetted area of these masks increase due to increased absorption. To illustrate this, we used special set of experiments with dyed droplets on white Mask C, and imaged the wetted area. As shown in Fig. \ref{fig:hydrophilic_all}d, the wetted area increases significantly with D\#. Such ability to spread the absorbed volume to the surrounding fibres, allows these masks to absorb more volume without getting fully saturated.    

Thus, the primary difference, in the context of droplet impact on dry vs. wet hydrophilic (such as cloth) mask is that, for the latter, the porous matrix already contains liquid. In the analysis for the critical condition for penetration of dry masks as shown in Eq. \ref{eq:Re_dry}, \citet{Sahu_porous_2012} evaluated 
the number of active pores participating in penetration as $N_d\sim(D/\epsilon)^3$, since only the pores under the droplet will be active (shown in Fig. \ref{fig:hydrophilic_all}c). However, when wet, the neighboring pores which already contains liquid will also be active (shown in Fig. \ref{fig:hydrophilic_all}c), and as such, the active number of pores for wet mask, will be given by, $N_w\sim(D_w/\epsilon)^3$, where $D_w$ is the mean diameter of the wetted region (shown in Fig. \ref{fig:hydrophilic_all}c). Substituting this in the analysis, we find that for wet hydrophilic mask, the penetration criteria will be  
\begin{equation}
Re(\epsilon/t_m)(D/D_w)^3\gg 1   
\label{eq:Re_wet}
\end{equation}
Furthermore, assuming the absorbed volume is homogeneously spread through the matrix of the wetted region, we can also write, $\widetilde{V}_a\sim D^2_wt_m \phi$, where the accumulated absorbed volume after $n$ impacts is $\widetilde{V}_a=\sum\limits_{i=1}^n (V_a)_i$, the absorbed volume during $i^{th}$ impact is $(V_a)_i$ and absorptivity of mask material, defined as volume of liquid absorbed per unit volume of the mask, is $\phi$. Recognizing that Eq. \ref{eq:Re_wet} needs to satisfied for penetration to occur, we can obtain the scaling relation for the critical velocity 
\begin{equation}
U_{cr}\sim B_0 ({\widetilde{V}^*_a})^{3/2}
\label{eq:scaling_hp}
\end{equation}
Here, normalized cumulative absorbed volume is $\widetilde{V}^*_a=\widetilde{V}_a/V$ and the initial droplet volume is $V\sim D^3$. The factor $B_0=\mu_l D^{3/2}/(\rho_l \epsilon^2t^{1/2}_m\phi^{3/2})$ is fixed for a given mask, liquid and droplet size. 
From Eq. \ref{eq:scaling_hp}, we see that for wet hydrophilic masks, the penetration process is strongly related to the amount of liquid absorbed by the mask matrix, not the number of impacted droplet, D\#.

\begin{figure}[]
\centering
\includegraphics[width=0.70\linewidth]{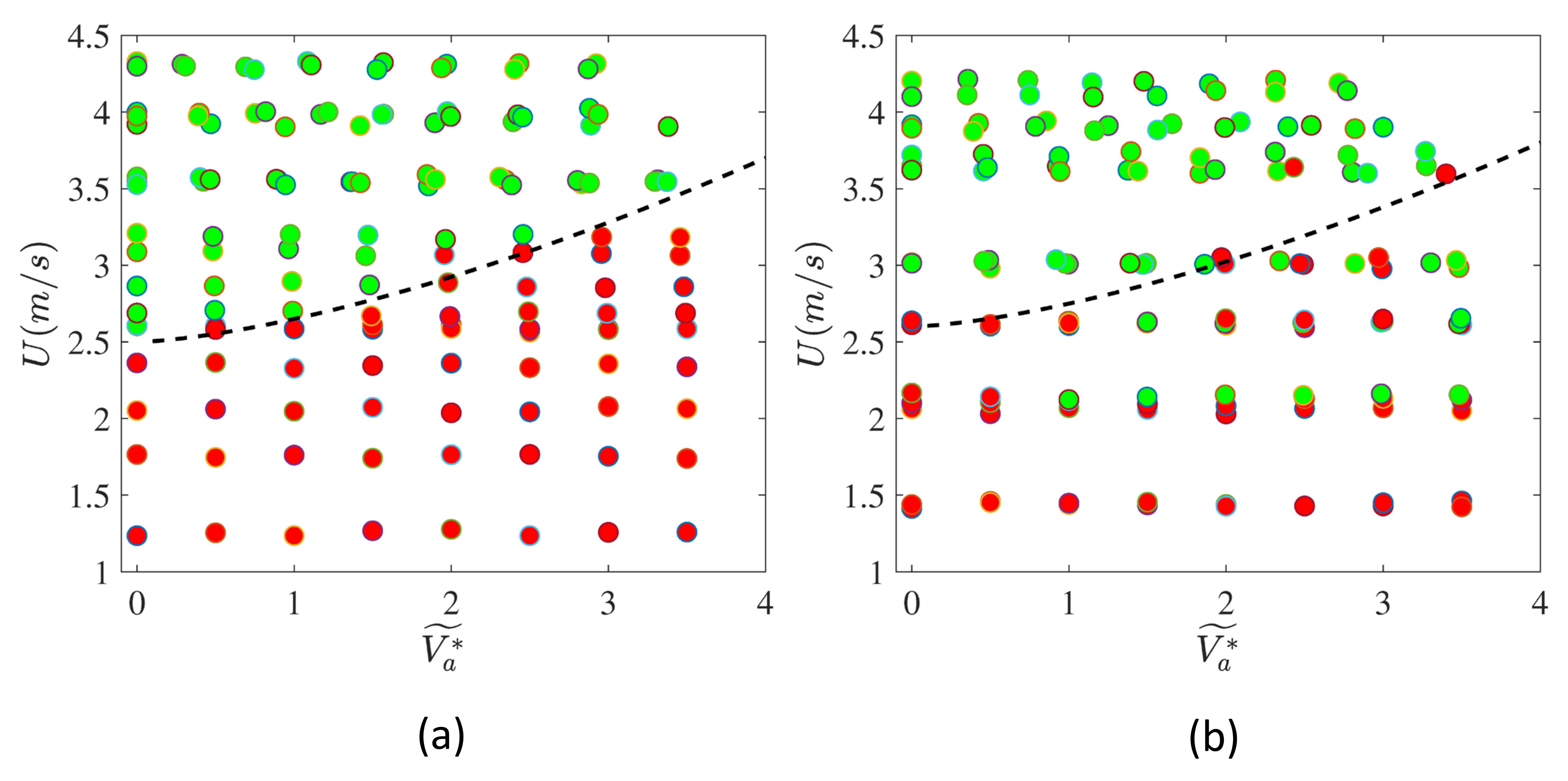}
\caption{Modified Regime diagram for hydrophilic masks as function of impact velocity ($U$) and normalized cumulative absorbed volume ($\widetilde{V^*_a}$). (a) Mask B and (b) Mask C. Green dots: penetration; Red dots: no penetration. The dashed line shows scaling:  $U_{cr}=U_0+B_a(\widetilde{V^*_a})^{3/2}$.}
\label{fig:Vabs_regime}
\end{figure}

In Fig. \ref{fig:Vabs_regime}a, we plotted the modified regime diagram for Mask B. Here, we represented each impact condition and outcome as function of normalized absorbed liquid volume before the impact, $\widetilde{V^*_a}$ and impact velocity, $U$. We note that, for certain value of $\widetilde{V^*_a}$ penetration occurs only when $U$ is greater than a critical value. This critical value, $U_{cr}$, identified by the boundary red (no penetration) and green (penetration) points, also increases with $\widetilde{V_a^*}$. Finally, to check the proposed scaling, we added a reference line, $U_{cr}=U_0+B_a(\widetilde{V^*_a})^{3/2}$, which shows a good agreement with the data. We note that $\widetilde{V_p}$ approaches zero for the first impact (D1), and as such, $U_0$ represents the critical condition for dry mask ($2.5 m/s$ for Mask B). We found that the scaling constant $B_a=0.15 m/s$ gives the best fit.  
We have also plotted the modified regime map for Mask C in Fig. \ref{fig:Vabs_regime}b. We notice that the transition between penetration vs no-penetration occurs for over a wider range as noted by the occurrences of both possible outcomes (red and green dots) at similar conditions. Nevertheless, if $U_{cr}$ is defined as the minimum velocity required for exclusively penetration outcome to occur for a given ${\widetilde{V}^*_a}$, we see that the proposed scaling (shown by dashed line) with $U_0=2.6 m/s$ can capture the transition behavior reasonably well.
It is to be noted that, the presented scaling does not consider any inhomogeneity in mask matrix or small variation in impact location between subsequent impacts, which are difficult to avoid in experiments. 

{In summary, we observe that the critical impact velocity required for the droplet to penetrate the wet mask, depends on the reflected volume for hydrophobic masks and the absorbed volume for hydrophilic masks. For both the masks, penetration becomes more difficult with increasing wetness, however, the underlying physics is different. For hydrophobic masks, there is negligible absorption. Thus, increased wetting results in larger reflected volume, which accumulates as beads on the mask surface. Since, the subsequently impacted droplet loses part of its kinetic energy to overcome the resistance provided by these water beads, penetration becomes weaker. On the other hand, due to high affinity to water of the  hydrophilic masks, the absorbed volume spreads homogeneously within the mask fibres. The porous matrix, thus, becomes filled with liquid, and the droplets during subsequent impacts will require to displace larger volume of liquid to penetrate the mask. Due to this additional resistance from absorbed liquid, the penetration becomes weaker.}


\subsection{Diameter of Daughter Droplets}

{The impact of a droplet on the mask matrix results in formation of ligaments and multiple daughter droplets. The disintegration of these ligaments and the secondary atomization results in generation of a large number of smaller droplets.} From high speed images, we have evaluated the size distribution of these droplets generated after penetration and secondary atomization. {We note that the droplet size distribution may vary with the downstream distance due to possible coalescence and subsequent breakups}\cite{kooij2019sprays}{. To be consistent, for all experiments, we measured the diameter distribution at a distance of 6 $mm$ from the masks.}  
For a given mask, significant differences in the Probability Density Functions (PDFs) among various impact numbers (i.e D1 vs D2 vs D3 etc) at similar impact velocities were not observed (see Supplementary Materials for example). Figure \ref{fig:pdf_v} shows the PDFs of daughter droplet diameters ($D_p$) for three different masks at various impact velocities (symbols with various colors) along with the overall distribution (black line). For hydrophobic masks (Mask A), we observe almost a Gaussian type distribution with most of the droplets ranging from $200\mu m <D_p< 600 \mu m$, and a peak around $D_p\approx 450 \mu m$. We do not see a significant difference in the PDFs among various $U$, as shown in Fig. \ref{fig:pdf_v}a. Here, we note that the penetration and secondary atomization is driven by the hydrodynamic focusing which was shown to be strongly dependent on pore size and impact velocity. In our previous study \cite{sharma2021secondary} with dry mask, we have shown that the PDF shifts towards a smaller droplet size as the velocity increased from $2m/s$ to to $10m/s$. Since, in the present experiments, the impact velocity range was relatively narrow ($2.5-4 m/s$), no discernible changes in PDF were observed. 

\citet{villermaux2007fragmentation} {showed that the variation in droplet sizes arising from breakup of ligaments can be represented by a gamma distribution which can be given as,}
\begin{equation}
    \Gamma(n,\overline{D}_p) = \frac{n^n}{\Gamma (n)}\overline{D}_{p}^{n-1}e^{-n \overline{D}_{p}},
\end{equation}
{where $\overline{D}_p=D_p/<D_p>$, $D_p$ is penetrated droplet diameter, $<D_p>$ the mean diameter of the secondary droplets and $n$ is the ligament corrugation parameter before destabilization. Highly corrugated ligaments corresponds to smaller values of $n$ (broader PDFs), while smooth ligaments have large values of $n$ (Narrower PDFs). For the hydrophobic mask (Mask A), we observe that $n=12$ (Fig. {\ref{fig:pdf_v}a}) provides a reasonable fit. Some discrepancies are expected as} \citet{villermaux2007fragmentation}
{proposed the Gamma function for shear-driven breakup of liquids surface, while in the present study the secondary droplets are generated through penetration of large droplets through highly inhomogeneous porous mask matrix. Furthermore, the wetness of the masks also affects the penetration and subsequent secondary atomization process.} 

{For hydrophilic masks (Mask B and C), the PDFs are shown in Fig. {\ref{fig:pdf_v}}b and c respectively. We see a relatively narrower droplet size distribution for Mask C (most droplets are in the range of $250\mu m <D_p < 500 \mu m$), compared to Mask B ($100\mu m <D_p < 600\mu m$) which is illustrated by their corresponding corrugation parameter values (Mask C: $n=20$, Mask B: $n=8$). The broader PDF in the Mask B can be attributed to the increased inhomogeneity in the mask matrix (see Supplementary Material for microscopic images). The effect of such inhomogeneity in pore distribution is also manifested through larger scatter in data among different impact velocities (Fig. {\ref{fig:pdf_v}b}).}

\begin{figure}[]
\centering
\includegraphics[width=0.8\linewidth]{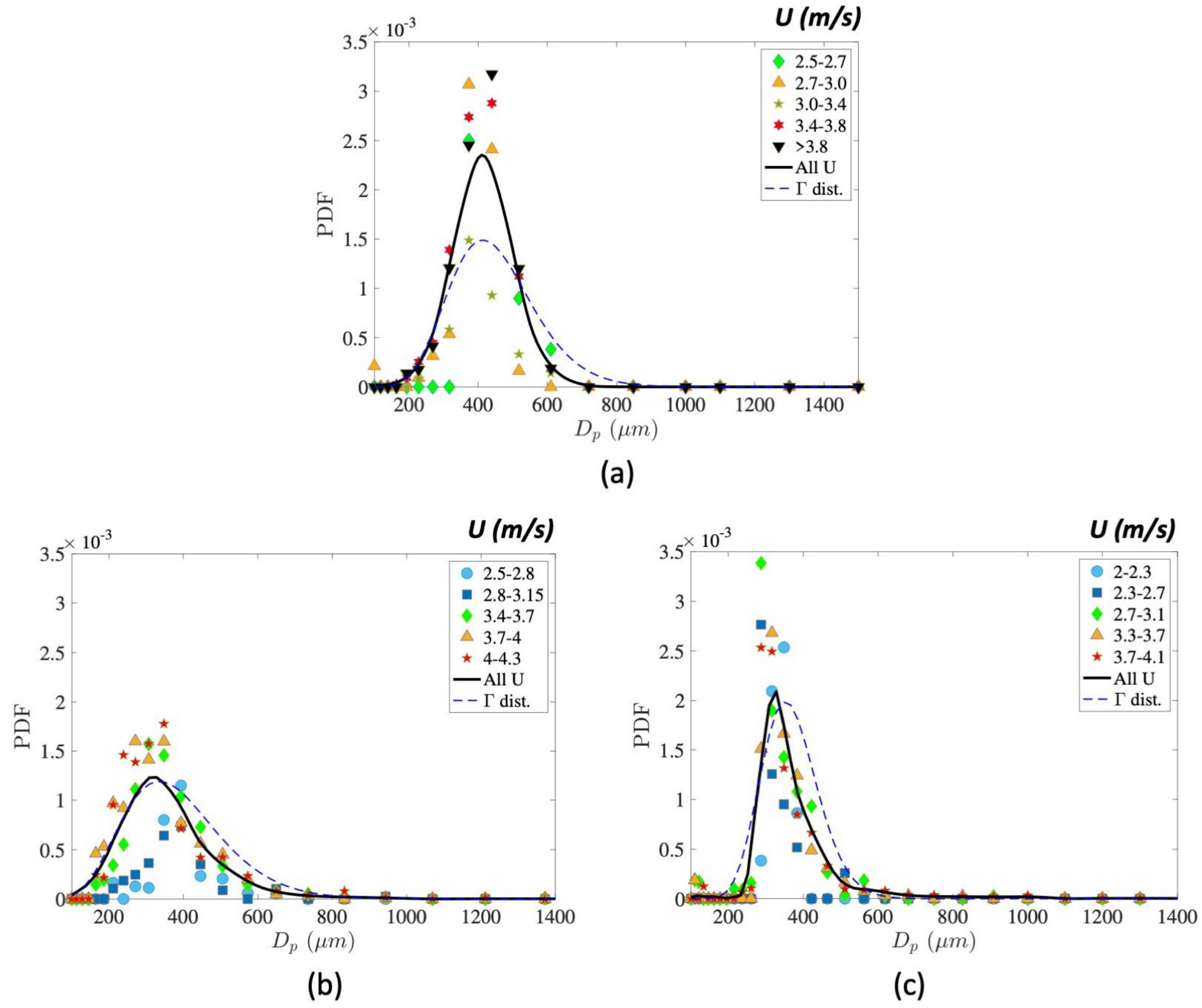}
\caption{{Probability Density Function (PDF) of diameter of secondary droplets generated during impact on (a) Mask A, (b) Mask B and (c) Mask C reported for different impact velocities, $U$.}}
\label{fig:pdf_v}
\end{figure}

\section{Discussion}
In this study, we investigated the penetration process of water droplets impacted on dry vs wet porous hydrophobic and hydrophilic masks. In dry condition, both types of masks show partial penetration and secondary atomization when impact velocity is beyond a critical value. For both masks, this critical value, however, increases as the masks get wet. Although there is a qualitative similarity between the hydrophobic and hydrophilic masks, the underlying mechanisms for delaying the penetration in wet masks are found to be different.

The porous matrix of hydrophobic masks cannot hold large quantity of water, and as such when the wetness increases during droplet impact and partial penetration, the remaining volume of water escapes the matrix to form water ``beads'' on the mask surface. The following droplets lands on these ``beads'', instead of directly impacting the porous matrix. The ``beads'', hence, provide additional resistance against penetration. Since the volume of the ``beads'' increase with the number of impacted droplets (and thus, wetness), the penetration becomes progressively more difficult. A scaling analysis based on energy conservation shows that the critical velocity, $U_{cr}$ scales as $U_{cr}\sim(V^*_r)^{1/2}$, where $V^*_r$ is the normalized volume of the ``bead'' before the impact. 

The porous matrix in hydrophilic masks, on the other hand, readily absorbs water and thus, as the mask gets wetter, the number of pores containing water increases. When a droplet impacts on such wet hydrophilic mask, it pushes liquid through a larger number of pores, compared to the number of active pores for impact on dry masks. This effectively, reduces the kinetic energy of liquid in each pore. As more droplets impact on the mask, and it absorbs more liquid, the number of active pores increases, hence progressively weakens the penetration process. A scaling analysis based on the energy balance for wetted hydrophilic masks resulted in the relation, $U_{cr}\sim ({\widetilde{V}^*_a})^{3/2}$, where $\widetilde{V}^*_a$ is the normalized cumulative absorbed volume in mask matrix.

We end this exposition by discussing the findings of the study in the context of protection offered by masks. First, we note that partial penetration and secondary atomization of large respiratory droplets through masks is not desirable.
In fact, the PDFs of the atomized droplets suggest that a large number of these droplets will have markedly smaller diameters compared to the initial droplet. Thus, if atomized through masks, large respiratory droplets may produce droplets smaller than $100 \mu m$. These ``aerosolized'' droplets can remain airborne for a long time, and thus can cause an increase in probability of pathogen transmission. However, for single layer dry masks, tested in this study, we observed the critical velocity required for partial penetration and secondary atomization to be more than $2 m/s$. Only strong respiratory events, such as cough, sneeze, can produce respiratory droplets with such velocity. For multi-layer masks, the critical velocity will be even higher, further reducing the probability of secondary atomization. 
Next, one might wonder, ``does prolonged use of masks reduce the protection"? For both (single layer) hydrophobic and hydrophilic masks, we found that the penetration becomes difficult when the mask is wet. While, the study does not account for some physical processes such as evaporation, condensation, bouncing of impacted droplets and dynamics of pathogens trapped in mask matrix, which may be important in some cases, the results of this study suggests that the wet masks do offer good capability in capturing exhaled respiratory droplets. The study, however, cannot assess the efficacy of the wet masks against inhaling aerosols or pathogen laden droplets. {We also acknowledge that the improper fitting of masks can lead to leakage of respiratory droplets and aerosols during both expiration and inhalation processes, minimizing the protection offered by the masks. Such effects are also not studied in this work.}  
More detailed study with targeted masks are needed for quantifying such efficacy.

\bibliographystyle{abbrvnat}
\bibliography{Reference.bib}

\end{document}